# Enhanced surface acceleration of fast electrons by using sub-wavelength grating targets

Guang-yue Hu[*], An-le Lei[†], Wen-tao Wang, Jing-wei Wang, Lin-gen Huang, Xin Wang, Yi Xu, Jian-sheng Liu, Bai-fei Shen, Wei Yu, Ru-xin Li, and Zhi-zhan Xu[‡]

(State Key Laboratory of High Field Laser Physics, Shanghai Institute of Optics and Fine Mechanics, Shanghai 201800, China)

**ABSTRACT**

Surface acceleration of fast electrons in intense laser-plasma interaction is improved by using sub-wavelength grating targets. The fast electron beam emitted along the target surface was enhanced by more than three times relative to that by using planar target. The total number of the fast electrons ejected from the front side of target was also increased by about one time. The method to enhance the surface acceleration of fast electron is effective for various targets with sub-wavelength structured surface, and can be applied widely in the cone-guided fast ignition, energetic ion acceleration, plasma device, and other high energy density physics experiments.

[*]gyhu@siom.ac.cn,  [†]lal@siom.ac.cn,  [‡]zzxu@mail.shcnc.ac.cn

**Text**

Fast ignition was proposed in the context of inertial fusion research to relax the strict symmetry requirements for implosion compression and to reduce the driver energy needed to form the ignition hot spot [1]. The core plasma is surrounded by hundred micron scale plasma. The fast ignition laser beam penetrates the long scale-length coronal plasma, and is absorbed and converted to fast electrons at around the critical density. The fast electrons are then transported into the core plasma for igniting. Complex physical process is involved in the propagation of heating laser in the long coronal plasma and thus its temporal and spatial stability are severely needed. So, an easy access of re-entrant cone was proposed to guide the intense heating laser beam close enough to the dense core plasma. The cone-guided fast ignition experiments have shown remarkable increase in the yield of thermal fusion neutrons [2]. The cone target has also been used in the field of ion acceleration and experiments of high energy density physics etc [19]. PIC simulations indicate that the cone targets can focus the laser energy and concentrate the fast electrons to the cone tip and enhance the energy coupling from the laser to fast electrons [3, 4]. Cone shape [5, 6], pre-plasma scale-length [7], polarization [8] and focusing properties of laser beam [6, 9] are investigated in order to increase the laser energy absorption and to concentrate more fast electrons to the cone tip.

The physical process of guiding and confining of fast electrons along the cone wall was explored individually by using planar target [10-15] or inverse cone target [16]. Surface acceleration of fast electrons was observed when the intense short pulse laser beam irradiated the planar target at large incident angle [13-15]. PIC simulations and analytical theory [10-12] show that a quasi-static surface magnetic field is induced by the fast electrons generated in the interaction via $J \times B$ heating or vacuum heating. A significant fraction of fast electrons injected into the plasma are deflected to the vacuum by the magnetic field, and pulled back again from the vacuum region by the charge separation field around the surface. Therefore, the fast electrons are confined around the surface and lead to the surface current which in turn enhances the surface magnetic field. It is similar to the surface current observed with wire targets [17-19].

The fast electrons confined inside the potential well of electrostatic potential and vector potential around the surface will oscillate at betatron frequency [11, 12]. Some of the oscillating fast electrons around the surface will be further accelerated by laser field through betatron acceleration mechanism if the resonance condition is matched [20]. Incident angle, pre-plasma scale-length, laser intensity, laser polarization, and interaction length (focus spot size) play important roles in the surface acceleration of fast electrons [10-15], which is similar to thata of cone target [5-9].

Due to the self-consistently enhanced quasi-static magnetic field and sheath electric field around the surface, the fraction of surface accelerated fast electrons can be enhanced by increasing the total number of the fast electrons generated in the laser-target interaction [10-15]. As shown in ref. [14], increasing the laser intensity is one straightforward method to improve the surface acceleration of fast electrons through enhancing the total number of the fast electron. There is an alternative method to enhance the total number of fast electrons, which does not need to increase the laser intensity. Targets with sub-wavelength structured surface such as foam, porous, nanometer-scale particles, clusters, nano-wires [21] have shown efficient enhancement of laser absorption and fast electron generation. Sub-wavelength grating targets, as the simplest material with structured surface of sub-wavelength (one dimension), also possesses near-complete laser absorption and enhanced x-ray emission induced by fast electrons [27]. Surface acceleration of fast electrons may prospectively be enhanced by using the targets with sub-wavelength structured surface.

In this article, we demonstrate that the targets with sub-wavelength structured surface are effective in enhancing the surface acceleration of fast electrons. By using sub-wavelength grating target, the fast electron beam emitted along the target surface was increased significantly, which is attributed to the enhanced laser absorption and fast electrons generation.

The experiments were carried out using a Ti:sapphire chirped pulse amplified laser system at a repetition rate of 10Hz. The laser beam (80mJ/60fs/795nm) was focused into a focus spot of 10 $\mu m$ (1/e$^2$ of the maximum intensity) with a f/4 off-axis parabolic mirror (20cm focus length). Planar aluminum slab and triangular ruled grating with aluminum (>1$\mu m$ thickness) coating on a glass substrate (Thorlabs, GR25-1800. 1800lines/mm, 555nm period, 279nm groove depth, and 26.7° blaze angle.) were used as the targets. As shown in Fig. 1, the *p*-polarized incident laser beam with central peak intensity of $\sim 0.5 \times 10^{18} W/cm^2$ irradiated the targets with $67.5^0 \pm 2^0$ incident angle, where ±2° is the accuracy of our target alignment. The contrast ratio of the incident laser pulse is better than 10$^{-8}$ at 50ps, 10$^{-7}$ at 7ps, and 10$^{-5}$ at 3ps before the main pulse, so pre-plasma formation before the main pulse is not significant. The emitted fast electrons (>400keV) were recorded by image plates (IP, Fujifilm BAS SR2025) placed in front of the target. Aluminum foils were placed before the image plates to block the x-ray and the electrons with lower energy.

Fig. 2 shows the measured angular distribution of fast electrons in the incident plane. The fast electrons generated by planar aluminum target mainly emit between the target normal direction and the incident laser beam, as shown in Fig. 2(b) and (c). Only a small fraction of fast electrons (10%-20%) are emitting between the laser specular and the target surface direction (Fig. 2(d)), which is the surface accelerated fast electrons [13-15, 23]. Fewer surface accelerated fast electrons relative to that reported previously [13, 14] should be due to the weaker laser intensity in our experiment. Weak laser intensity is unfavorable for the formation of the surface quasi-static electric field and magnetic field that cause the surface acceleration of fast electrons [13-15]. The fast electrons emitted between the target normal direction and the incident laser beam have also been reported elsewhere, but the generation mechanism is not clear yet [15, 22, 23].

By using grating targets, as shown in Fig. 2(a), (c), and (d), a large amount of fast electrons are emitting along the target surface, which is more than 3 times higher than that by using planar target. The fraction of surface accelerated fast electrons is increased from 10%-20% for planar target to about ~50% for grating target, which is close to that by using planar target with much higher laser intensity of $\sim 2\times 10^{18} W/cm^2$ [13-15]. The total numbers of the fast electrons emitted from the front side of target are also enhanced by about one time, as indicated in Fig. 2(d).

Two-dimension particle-in cell (2D PIC) simulations are performed to explain the experimental results. In the simulation, we take planar laser pulses which have normalized amplitude $a_0 = 1$, laser wavelength $\lambda_0 = 0.8\mu m$, pulse duration $15T_0$ [ $I(t) = \sin^2\left(\pi t/2\tau_0\right), \tau_0 = 15T_0$, $T_0$ is the laser period], focus spot $10\mu m$ [$I(r) = \exp(-2r/r_0), r_0 = 5\mu m$], and $67.5°$ incident angle. The geometric shape of the simulated grating targets is identical to that used in our experiment. The initial density of the target plasma distributed within $z \sim (5-7)\mu m$ is $n_e/n_c = 10$. The peak of the laser pulse reaches target surface at about $30T_0 - 31T_0$.

Fig. 3 (d) and (f) show the negative quasi-static (averaged in one cycle) electric field $<E_z>$ and magnetic field $<B_x>$ around the front surface of the grating target at $29T_0 - 30T_0$, which exhibit evident character of surface acceleration of fast electrons, as discussed in the case of planar target [10-12, 14]. Moreover, the amplitude of $<E_z>$ and $<B_x>$ around the surface of the grating target shown in Fig. 3 (d) and (f) are obviously stronger than that of the planar target in Fig. 3 (c) and (e), which indicating enhanced surface acceleration of fast electrons. The enhanced surface acceleration of fast electrons should be due to the improved laser absorption and fast electrons generation in the laser-grating interaction, as indicated by the increase of the sheath electric field $E_z$ around the rear surface of the grating target in Fig. 3 (d) and Fig. 4(b). Because the incident angle of laser beam departs far from the resonant angle, the improved laser absorption and fast electrons generation by using grating target should not be induced by surface plasmon resonance excitation [24-28]. However, the wavelength-scale gibbous cell of sub-wavelength grating target can also lead to very strong enhancements of the laser field and the laser absorption through Mie resonances and other mechanisms [21]. As shown in Fig. 4 (b) and (d), the electric fields $E_z$ and $E_y$ around the gibbous cells of grating target are increased significantly relative to that around the surface of the planar target in Fig. 4 (a) and (c). Similar results have been reported with other shapes of sub-wavelength microstructure [21, 27, 29]. This enhanced electric field around the sub-wavelength gibbous cells can improve the laser absorption and fast electrons generation, which has been confirmed in experiment with wavelength-scale spheres and foam etc [21]. Different from the surface plasmon resonance, which requires the laser beam must irradiate the periodic structure (such as grating target) with a certain resonant angle, the present local field enhancement is attributed to the wavelength-scale gibbous cells and is closely relavant to the size of the wavelength-scale microstructure [21], which is also observed in PIC simulation of grating target [27, 29]. The size and the depth of the grating can efficiently change the fast electron generation and the surface acceleration of fast electrons [29]. Periodic structure is not necessary for the present local field enhancement. For the materials with 3D wavelength-scale microstructure such as foam and porous etc, the irregular microstructure can easily satisfy the optimal size. Moreover, they can survive during the picoseconds interaction [21]. So the side wall of the reentrant cone used for fast ignition can be covered by low-density foam with suitable thickness to improve the fast electrons generation and guiding to the cone tip. For the experiments using short pulse laser, materials with nanometer-scale structured surface, such as nano-particles, clusters, nano-wire, velvet target etc [21], can also be used to enhance the surface acceleration of fast electrons.

It is worth pointing out that the periodic surface structure of grating target may be served as a slow wave device and help to the phase matching between the oscillating fast electrons in surface potential well with the laser electric field [11, 12]. More fast electrons with lower energy are resonantly accelerated to high energy at the stage of betatron acceleration.

In summery, surface acceleration of fast electron was enhanced significantly by using sub-wavelength grating targets. The improved surface accelerated fast electrons are beneficial from the increased laser energy absorption and fast electrons generation with the wavelength-scale structured surface of grating target. Materials with nanometer-scale or wavelength-scale structured surface such as foam, porous, nano-particles, and clusters etc, which are easier to fabricate and cheaper than grating targets, can probably be used to improve the surface acceleration of fast electrons, and are useful in the cone-guiding fast ignition, cone-attached ion acceleration, plasma devices [19], and other high energy density physics experiments.

**ACKNOWLEDGMENTS**

This work was supported by National Natural Science Foundation of China under Grant Nos. 10875158, 60921004, and 10775165, the Science and Technology Commission of Shanghai Municipality under Grant No. 08PJ14102, and National Basic Research Program of China (973 Program) under Grant No. 2006CB806000.

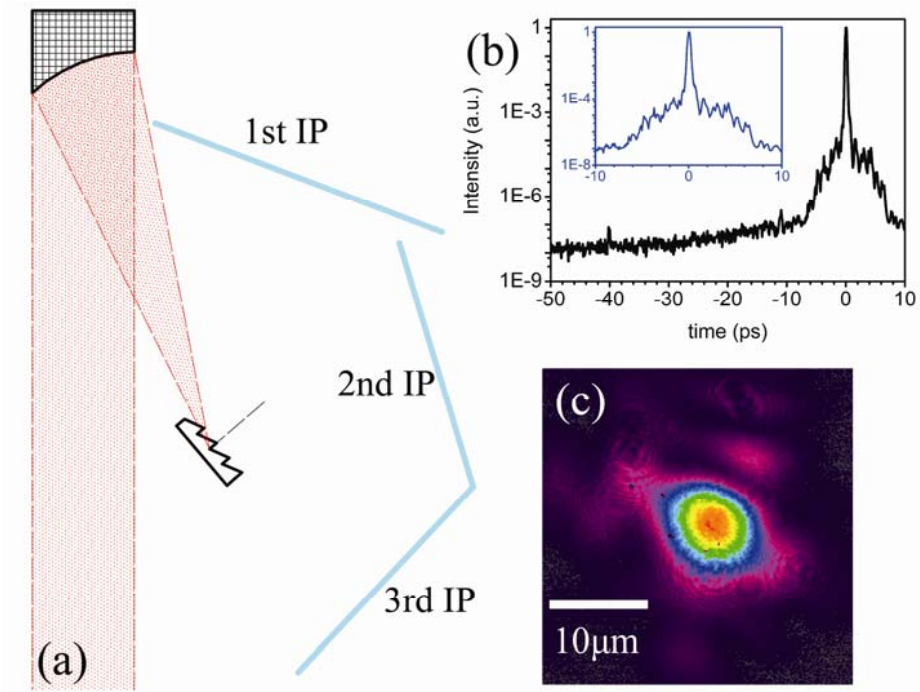

Fig. 1 (Color online) (a) Scheme of the experimental setup. IP is the image plate with 20cm height and 15cm (1st), 12.5cm (2nd), and 12.5cm (3rd) length. (b) Contrast ratio of the incident laser pulse measured by a third-order autocorrelation device. (c) Laser focus spot.

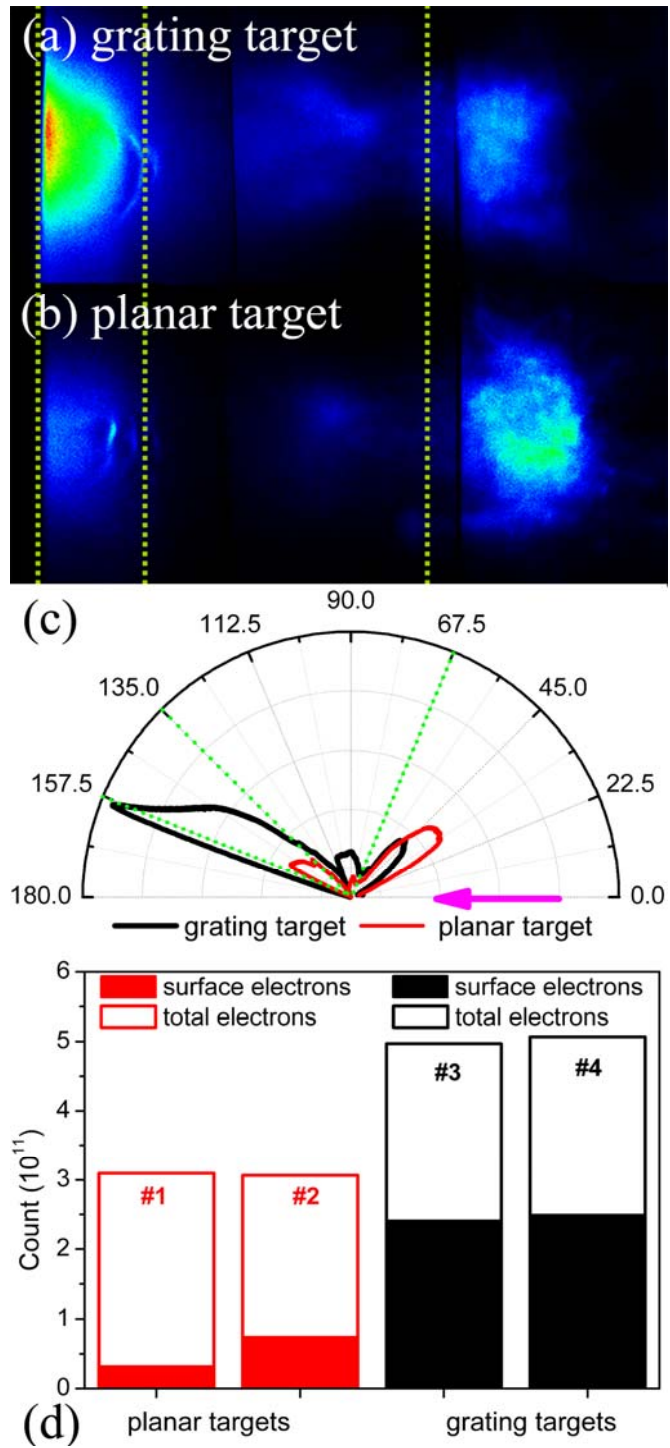

Fig. 2 (Color online) Fast electron emission distribution recorded on IPs (>400keV) generated by grating targets (a) and planar targets (b). The vertical lines represent the angles for target surface, laser specular, and target normal directions from left to right. (c) Angular distribution curves of fast electrons in the incident plane. The green dot lines, as thaose in (a) and (b), are respectively the angles of the target surface, laser specular, and target normal directions from left to right. Laser beam is labeled with the arrow. (d) The counts on IP of the surface accelerated fast electrons and the total fast electrons emitted toward the front side of target. #1 and #2 are the shots for planar target, #3 and #4 are the shots for grating target.

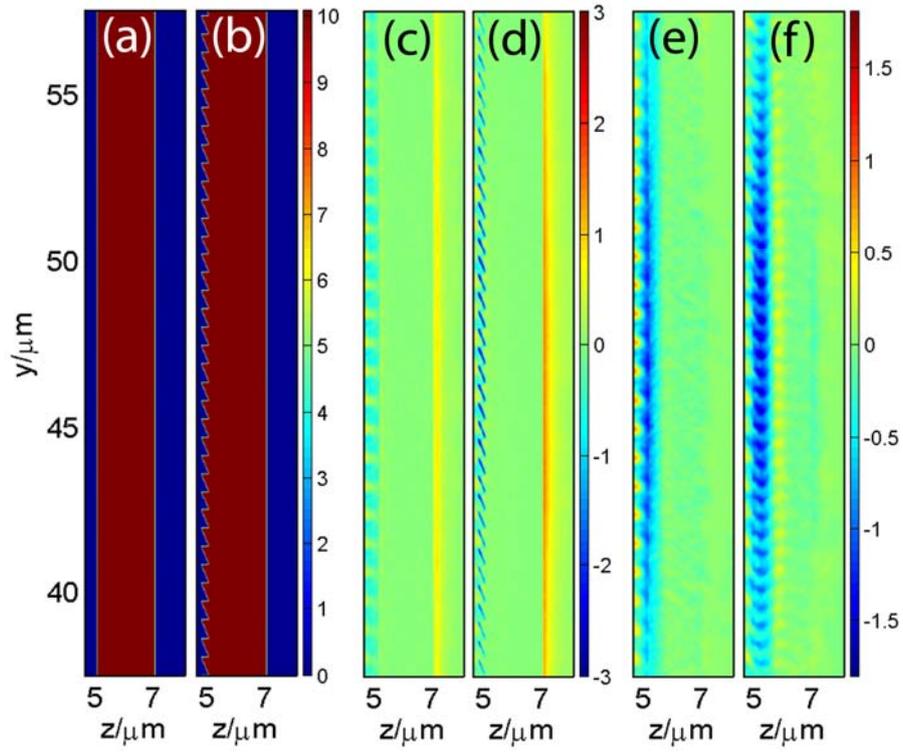

Fig. 3 (Color online) The spatial distributions of the initial electron density (a, b), the longitudinal static electric field $<E_z>$ (c, d), and the static magnetic field $<B_x>$ (e, f) cycle-averaged within $29T_0 - 30T_0$. (a, c, d) are for the planar target, (b, d, f) are for the grating target.

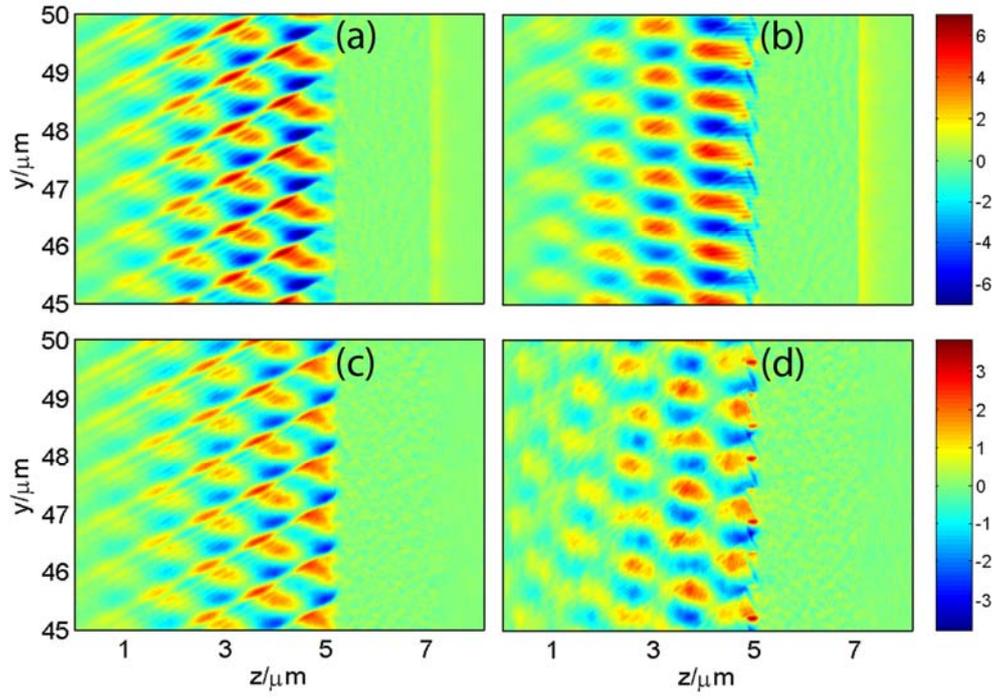

Fig. 4 (Color online) The spatial distributions of longitudinal electric field $E_z$ (a, b), and transverse electric field $E_y$ (c, d) at $29T_0$. (a, c) are for the planar target, (b, d) are for the grating target.